\documentclass[12pt]{article}
\usepackage{graphicx}
\usepackage{epsf}

\makeatletter
\@addtoreset{equation}{section}
\makeatother

\let\la=\label  
  
 \def\bd{\begin{document}} \def\ed{\end{document}}
\def\ds{\documentstyle} \let\fr=\frac \let\bl=\bigl \let\br=\bigr
\let\Br=\Bigr \let\Bl=\Bigl
\let\bm=\bibitem
\let\na=\nabla
\let\pa=\partial \let\ov=\overline
\newcommand{\be}{\begin{equation}}
\newcommand{\ee}{\end{equation}}
\def\ba{\begin{array}}
\def\ea{\end{array}}
\newcommand{\ho}[1]{$\, ^{#1}$}
\newcommand{\hoch}[1]{$\, ^{#1}$}
\newcommand{\bea}{\begin{eqnarray}}
\newcommand{\eea}{\end{eqnarray}}
\newcommand{\ra}{\rightarrow}
\newcommand{\lra}{\longrightarrow}
\newcommand{\Lra}{\Leftrightarrow}
\newcommand{\ap}{\alpha^\prime}
\newcommand{\bp}{\tilde \beta^\prime}
\newcommand{\tr}{{\rm tr} }
\newcommand{\Tr}{{\rm Tr} }
\newcommand{\NP}{Nucl. Phys. }
\ifx\ltimes\undefined
\newcommand{\ltimes}{{\kern3pt\hbox{\vrule width 0.4pt height 5.30pt depth
.0pt}\kern-1.76pt\times\kern1pt}}
\fi

\newcommand{\tamphys}{\it $^\dag$ The Blackett Laboratory, Imperial College London,\\ 
Prince Consort Road, London SW7 2AZ\\
~~\\
$^\ddag$Physics Department,Theory Unit, CERN, CH1211, Geneva23, 
Switzerland\\
and\\ 
Department of Physics \& Astronomy, Universuty of California,Los Angeles, 
USA\\
and\\ INFN-Laboratori Nazionale di Frascati, Via E. Fermi 40, 00044 Frascati, Italy}
\newcommand{\auth}{M. J. Duff \footnote{m.duff@imperial.ac.uk}$^\dag$ and S. 
Ferrara\footnote{Sergio.Ferrara@cern.ch}$^{\ddag}$}

\begin{document}
\begin{flushright}
\hfill{Imperial/TP/2007/mjd/1}\\
\hfill{CERN-PH-TH/2007-78}\\
\hfill{UCLA/07/TEP/7}\\
\end{flushright}
\vspace{10pt}

\begin{center}
{ \large {\bf $E_{6}$ and the bipartite entanglement of three qutrits}}

\vspace{20pt}

\auth

\vspace{10pt}

{\tamphys}

\vspace{20pt}

\underline{ABSTRACT}

\end{center}

Recent investigations have established an analogy between the entropy of 
four-dimensional supersymmetric black holes in string theory 
and entanglement in quantum information theory. 
Examples include: (1) $N=2$ STU black holes and the tripartite entanglement 
of three qubits (2-state systems), where the common symmetry is 
$[SL(2)]^{3}$ and (2) $N=8$ black holes and the tripartite 
entanglement of seven qubits where the common symmetry is $E_{7} 
\supset [SL(2)]^{7}$. Here 
we present another example: $N=8$ black holes (or black strings) in five dimensions and 
the bipartite entanglement of three qutrits (3-state systems), where the common 
symmetry is $E_{6} \supset [SL(3)]^{3}$. Both the black hole (or 
black string) entropy and the 
entanglement measure are provided by the Cartan cubic $E_{6}$ invariant. Similar analogies 
exist for ``magic'' $N=2$ supergravity black holes in both four and five dimensions.

\vfill
\leftline{}

\newpage

\tableofcontents

\newpage


\section{$D=4$ black holes and qubits}

It sometimes happens that two very different areas of theoretical physics share 
the same mathematics.  This may eventually lead to the realisation that they are, in fact,
dual descriptions of the same physical phenomena, or it may not.  Either way, it 
frequently leads to new insights in both areas.  Recent papers 
\cite{Duff:2006uz,Kallosh:2006zs,Levay:2006kf,Duff:2006ue,Levay:2006pt,Duff:2006rf}
have established an analogy between the entropy of certain four-dimensional supersymmetric 
black holes in string theory and entanglement measures in quantum information theory. In this 
paper we extend the analogy from four dimensions to five which also involves going 
from two-state systems (qubits) to three-state systems (qutrits).

We begin by recalling the four-dimensional examples:

\subsection{$N=2$ black holes and the tripartite entanglement of three qubits} 
\la{N=2}

The three qubit system (Alice, Bob, Charlie) is described by the state 
\begin{equation}
|\Psi\rangle = a_{ABC}|ABC\rangle
\end{equation}
where $A=0,1$, so the Hilbert space has dimension $2^{3}=8$.  
The complex numbers $a_{ABC}$ transforms as a $(2,2,2)$ under 
$SL(2, C)_{A} \times SL(2, C)_{B} \times SL(2, C)_{C}$. The tripartite 
entanglement is measured by the {\it 3-tangle} \cite{Coffman,Miyake}
\be
\tau_{3}(ABC)=4 |{\rm Det}~a_{ABC}|.
\ee
where ${\rm Det}~a_{ABC}$ is Cayley's hyperdeterminant \cite{Cayley}. 
\be
{\rm Det}~a=-\frac{1}{2}
\epsilon^{A_{1}A_{2}}\epsilon^{B_{1}B_{2}}\epsilon^{A_{3}A_{4}}\epsilon^{B_{3}B_{4}}
\epsilon^{C_{1}C_{4}}\epsilon^{C_{2}C_{3}}
{a}_{A_{1}B_{1}C_{1}}{a}_{A_{2}B_{2}C_{2}}{a}_{A_{3}B_{3}C_{3}}{a}_{A_{4}B_{4}C_{4}}
\la{Cayley}
\ee
The hyperdeterminant is invariant under $SL(2)_{A}\times SL(2)_{B}\times SL(2)_{C}$ and 
under a triality that interchanges $A$, $B$ and $C$. 

In the context of stringy black 
holes the 8 $a_{ABC}$ are the 4 electric and 4 
magnetic charges of the $N=2$ STU black hole \cite{Duff:1995sm} and hence take 
on real (integer) values. The $STU$ model corresponds to $N=2$ supergravity coupled to 
three vector multiplets, where the symmetry is $[SL(2,Z)]^{3}$. The 
Bekenstein-Hawking entropy of the black hole, $S$, was first calculated in 
\cite{Behrndt:1996hu}. The connection to quantum information theory 
arises by noting \cite{Duff:2006uz} that it can also be expressed in terms of 
Cayley's hyperdeterminant
\be
 S= \pi \sqrt{|{\rm Det}~a_{ABC}|}.
\ee  
One can then establish a dictionary between the classification of 
various entangled 
states (separable A-B-C; bipartite entangled A-BC, B-CA, C-AB; 
tripartite entangled W; tripartite entangled GHZ) and the 
classfication of various ``small'' and ``large'' BPS and non-BPS black 
holes \cite{Duff:2006uz,Kallosh:2006zs,Levay:2006kf,Duff:2006ue,Levay:2006pt,Duff:2006rf}.
For example, the GHZ state 
\cite{Greenberger}
\be
|\Psi\rangle \sim |111\rangle + |000\rangle
\ee
with ${\rm Det}~a_{ABC} \geq 0$ corresponds to a large non-BPS 2-charge 
black hole; the W-state
\be
|\Psi\rangle \sim |100\rangle+|010\rangle+|001\rangle
\ee
with ${\rm Det}~a_{ABC} = 0$ corresponds to a small-BPS 3-charge 
black hole; the GHZ-state
\be
|\Psi\rangle= -|000\rangle +|011\rangle+|101\rangle +|110\rangle
\la{GHZ}
\ee
corresponds to a large BPS 4-charge black hole.

\subsection{$N=2$ black holes and the bipartite entanglement of two qubits} 
\la{N=4,D=4}

An even simpler example \cite{Kallosh:2006zs} is provided by the two qubit 
system (Alice and Bob) described by the state
\be
|\Psi\rangle = a_{AB}|AB\rangle 
\ee
where $A=0,1$, and the Hilbert space has dimension $2^{2}=4$.  The $a_{AB}$ transforms 
as a $(2,2)$ under $SL(2, C)_{A} \times SL(2, C)_{B}$. The entanglement is
measured by the {\it 2-tangle}
\be
\tau_{2}(AB)=C^{2}(AB)
\ee
where
\be
C(AB)=2~|{\rm det}~a_{AB}|
\ee
is the {\it concurrence}.  The determinant is invariant under $SL(2, C)_{A} \times SL(2, C)_{B}$
and under a duality that interchanges $A$ and $B$. Here it is 
sufficient to look at $N=2$ supergravity coupled to just one vector 
multiplet and the 4 $a_{AB}$ are the 2 electric and 2 magnetic charges 
of the axion-dilaton black hole with entropy
\be
 S= \pi|{\rm det}~a_{AB}|
\ee  
For example, the Bell state
\be
|\Psi\rangle \sim |11\rangle + |00\rangle
\ee
with ${\rm det}~a_{AB} \geq 0$ corresponds to a large non-BPS 2-charge black hole.

\subsection{$N=8$ black holes and the tripartite entanglement of seven qubits} 
\la{N=8,D=4}

We recall that in the case of $D=4,N=8$ supergravity, the the 28 
electric and 28 magnetic charges  belong to the $56$ of $E_{7(7)}$. The black hole entropy is 
\cite{Kallosh:1996uy,Ferrara:1997uz}

\be
S=\pi\sqrt{|J_{4}|}
\la{3/7entropy}
\ee
where $J_{4}$ is Cartan's quartic $E_{7}$ invariant \cite{Cartan,Cremmer:1979up}. It may be 
written 
\begin{eqnarray}
J_{4} &=&  P^{ij}  Q_{jk}  P^{kl}  Q_{li} - {\textstyle{1\over 4}} P^{ij}  Q_{ij}
P^{kl}  Q_{kl} \nonumber\\
 \nonumber\\
&+& {\textstyle{1\over 96 }}   \Bigl(\epsilon^{ijklmnop} \,  Q_{ij}  Q_{kl}
Q_{mn}  Q_{op}  + \epsilon_{ijklmnop} \,   P^{ij}   P^{kl}   P^{mn}   
P^{op}
\Bigr) \ .
\label{Car}\end{eqnarray}
where  $P^{ij}$ and $Q_{jk}$ are $8 \times 8$ antisymmetric matrices.

The qubit interpretation \cite{Duff:2006ue} relies on the decomposition
\be
E_{7}(C) \supset [SL(2,C)]^{7}
\ee
under which
\[
56 
\rightarrow
\]
\[
~(2,2,1,2,1,1,1)  
\]
\[ 
+(1,2,2,1,2,1,1)
\]
\[
+(1,1,2,2,1,2,1)      
\]
\[
+(1,1,1,2,2,1,2)              
\]
\[
+(2,1,1,1,2,2,1)                                            
\]
\[ 
+(1,2,1,1,1,2,2) 
\]
\be
 +(2,1,2,1,1,1,2)
\la{decomp}
\ee
suggesting the tripartite entanglement of seven qubits (Alice, Bob, Charlie, Daisy, Emma, Fred and 
George) described by the state.
\[
|\Psi\rangle = 
\]
\[
                a_{ABD}|ABD\rangle
                \]
                \[
               +b_{BCE}|BCE\rangle
               \]
               \[
               +c_{CDF}|CDF\rangle
               \]  
               \[
               +d_{DEG}|DEG\rangle
               \]
               \[
               +e_{EFA}|EFA\rangle
               \]
               \[
               +f_{FGB}|FGB\rangle
               \]
\be
               +g_{GAC}|GAC\rangle
			   \la{3/7psi}
\ee
where $A=0,1$, so the Hilbert space has dimension $7.2^{3}=56$. The 
$a,b,c,d,e,f,g$ transform as a $56$ of $E_{7}(C)$.  The entanglement may be represented by a 
heptagon where the vertices A,B,C,D,E,F,G represent the seven qubits 
and the seven triangles ABD, BCE, CDF, DEG, EFA, FGB, GAC represent 
the tripartite entanglement. See Figure 1.
\begin{figure}[ht]
\begin{center}
\epsfysize=7cm\epsfbox{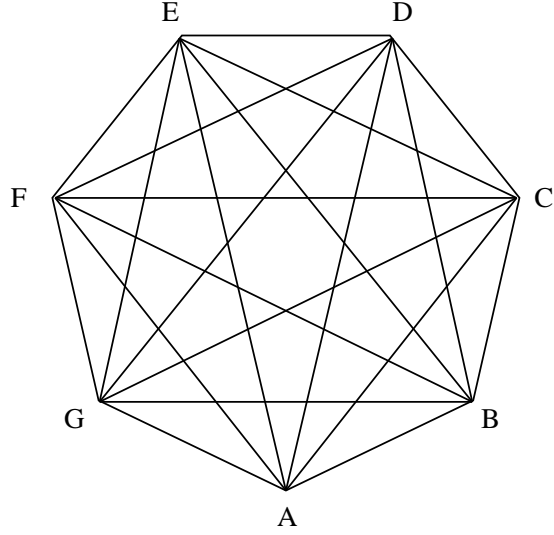}
\end{center}
\caption{The $E_7$ entanglement diagram. Each of the seven vertices A,B,C,D,E,F,G 
represents a qubit and each of the seven triangles ABD, BCE, CDF, DEG, 
EFA, FGB, GAC describes a tripartite entanglement.  }
\end{figure} 
Alternatively, we can use the Fano plane. See Figure 2.  The Fano plane also 
corresponds to the multiplication table of the  
octonions\footnote{Not the ``split'' octonions as was incorrectly stated in 
the published version of \cite{Duff:2006ue}.} 
\begin{figure}[ht]
\begin{center}
\epsfysize=7cm\epsfbox{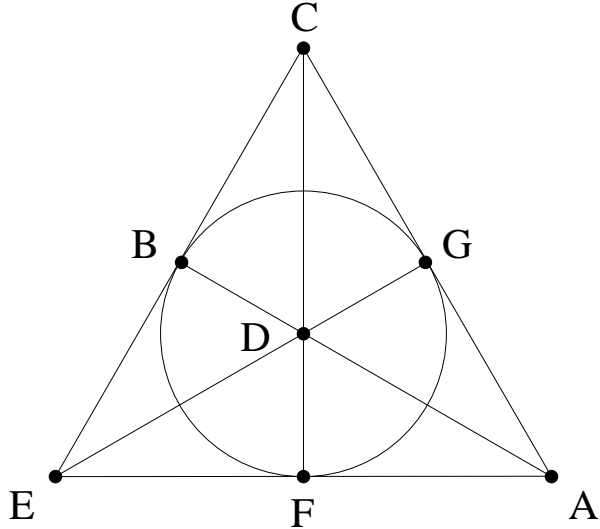}
\end{center}
\caption{The Fano plane has seven points, representing the seven 
qubits, and seven lines (the circle 
counts as a line) with three points on every line, representing the 
tripartite entanglement, and three lines through every point.}
\end{figure}

The measure of the tripartite entanglement of the seven qubits is provided by the {\it 3-tangle}
\be
\tau_{3}(ABCDEFG) = 4 |J_{4}|
\la{3/7tangle}
\ee
with
\[
J_{4}\sim a^{4}+b^{4}+c^{4}+d^{4}+e^{4}+f^{4}+g^{4}+
\]
\[
2[a^{2}b^{2}+b^{2}c^{2}+c^{2}d^{2}+d^{2}e^{2}+e^{2}f^{2}+f^{2}g^{2}+g^{2}a^{2}+
\]
\[
  a^{2}c^{2}+b^{2}d^{2}+c^{2}e^{2}+d^{2}f^{2}+e^{2}g^{2}+f^{2}a^{2}+g^{2}b^{2}+
\]
\[
 a^{2}d^{2}+b^{2}e^{2}+c^{2}f^{2}+d^{2}g^{2}+e^{2}a^{2}+f^{2}b^{2}+g^{2}c^{2}]
\]
\be  
+8[bcdf+cdeg+defa+efgb+fgac+gabd+abce]
\la{564} 
\ee    
where products like
\[
a^{4}= (ABD)(ABD)(ABD)(ABD)
\]
\be
=\epsilon^{A_{1}A_{2}}\epsilon^{B_{1}B_{2}}\epsilon^{D_{1}D_{4}}\epsilon^{A_{3}A_{4}}\epsilon^{B_{3}B_{4}}\epsilon^{D_{2}D _{3}}
{a}_{A_{1}B_{1}D_{1}}{a}_{A_{2}B_{2}D_{2}}{a}_{A_{3}B_{3}D_{3}}{a}_{A_{4}B_{4}D_{4}}
\ee
exclude four individuals (here Charlie, Emma, Fred and George), products like
\[
a^{2}b^{2}=(ABD)(ABD)(FGB)(FGB)
\]
\be
=\epsilon^{A_{1}A_{2}}\epsilon^{B_{1}B_{3}}\epsilon^{D_{1}D_{2}}\epsilon^{F_{3}F_{4}}\epsilon^{G_{3}G_{4}}\epsilon^{B_{2}B_{4}}
{a}_{A_{1}B_{1}D_{1}}{a}_{A_{2}B_{2}D_{2}}{b}_{F_{3}G_{3}B_{3}}{b}_{F_{4}G_{4}B_{4}}
\ee
exclude two individuals (here Charlie and Emma), and products like
\[
abce= (ABD)(BCE)(CDF)(EFA)
\]
\be
=\epsilon^{A_{1}A_{4}}\epsilon^{B_{1}B_{2}}\epsilon^{C_{2}C_{3}}\epsilon^{D_{1}D_{3}}\epsilon^{E_{2}E_{4}}\epsilon^{F_{3}F_{4}}
{a}_{A_{1}B_{1}D_{1}}{b}_{B_{2}C_{2}E_{2}}{c}_{C_{3}D_{3}F_{3}}{e}_{E_{4}F_{4}A_{4}}
\ee
exclude one individual (here George)\footnote{This corrects the 
corresponding equation in the published version of \cite{Duff:2006ue} which had the wrong 
index contraction.}. 

Once again large non-BPS, small BPS and large BPS black holes correspond to 
states with $J_{4} > 0$, $J_{4} = 0$ and  $J_{4} < 0$, respectively.

\subsection{Magic supergravities in $D=4$}
\la{magic4}

The black holes described by Cayley's hyperdeterminant are those of 
$N=2$ supergravity coupled to three vector multiplets, where the 
symmetry is $[SL(2,Z)]^{3}$.  In \cite{Duff:2006ue} the following 
four-dimensional generalizations were considered:

1) $N=2$ supergravity coupled to $l$ vector multiplets where the 
symmetry is $SL(2,Z) \times SO(l-1,2,Z)$ and the black holes carry charges 
belonging to the $(2,l+1)$ representation ($l+1$ electric plus $l+1$ magnetic).

2) $N=4$ supergravity coupled to $m$ vector multiplets where the 
symmetry is $SL(2,Z) \times SO(6,m,Z)$ where the black holes carry 
charges belonging to the $(2,6+m)$ representation ($m+6$ electric plus 
$m+6$ magnetic). 

3) $N=8$ supergravity where the symmetry is the non-compact 
exceptional group $E_{7(7)}(Z)$ and the black 
holes carry charges belonging to the fundamental $56$-dimensional 
representation (28 electric plus 28 magnetic).

In all three case there exist quartic invariants akin to Cayley's 
hyperdeterminant whose square root yields the corresponding black hole 
entropy. In \cite{Duff:2006ue} we succeeded in giving a quantum theoretic 
interpretation in the $N=8$ case together with its truncations to 
$N=4$ (with $m=6$) and $N=2$ (with $l=3$, the case we already knew 
\cite{Duff:2006uz}).

However, as suggested by Levay \cite{Levay:2006pt}, one might also 
consider the ``magic'' supergravities 
\cite{Gunaydin:1984ak,Gunaydin:1983bi,Gunaydin:1983rk}. These correspond to the R, C, 
H, O (real, complex, quaternionic and octonionic) $N=2,D=4$ 
supergravity coupled to $6,9,15$ and $27$ vector multiplets with 
symmetries $Sp(6,Z),SU(3,3),SO^{*}(12)$ and $E_{7(-25)}$, respectively.
Once again, as has been shown just recently \cite{Ferrara:2006yb}, in all cases 
there are quartic invariants whose square root yields the corresponding black hole 
entropy. 

Here we demonstrate that the black-hole/qubit correspondence does 
indeed continue to hold for magic supergravities. The crucial observation is that, although 
the black hole charges $a_{ABC}$ are real (integer) numbers and the 
entropy (\ref{3/7entropy}) is invariant under $E_{7}(7)(Z)$,  the coefficients 
$a_{ABC}$ that appear in the qutrit state (\ref{3/7psi}) are complex. So the three 
tangle (\ref{3/7tangle}) is invariant under $E_{7}(C)$ which contains 
both $E_{7(7)}(Z)$ and $E_{7(-25)}(Z)$ as subgroups.  To find a 
supergravity correspondence therefore, we could equally well have chosen  
the magic octonionic $N=2$ supergravity rather than the conventional 
$N=8$ supergravity. The fact that 
\be
E_{7(7)}(Z) \supset [SL(2)(Z)]^{7}
\ee
but
\be
E_{7(-25)}(Z) \not  \supset [SL(2)(Z)]^{7} 
\ee
is irrelevant. All that matters is that
\be
E_{7}(C) \supset [SL(2)(C)]^{7}
\ee
The same argument holds for the magic real, complex and quaternionic 
$N=2$ supergravities which are, in any case truncations of $N=8$ (in contrast 
to the octonionic) . 

Having made this observation, one may then revisit the conventional 
$N=2$ and $N=4$  cases (1) and (2) above.  When we looked at the seven 
qubit subsector $E_7(C) \supset SL(2,C)  \times SO(12, C)$, we gave an 
$N=4$ supergravity interpretationÊwith symmetry $SL(2,R) \times SO(6,6)$ 
\cite{Duff:2006ue}, but we could equally 
have given an interpretation in terms of $N=2$ supergravity coupled to 
$11$ vector multiplets with 
symmetry $SL(2,R) \times SO(10,2)$.

Moreover, $SO(l-1,2)$ is contained in $SO(l+1,C)$ and $SO(6,m)$ is contained in $SO(12+m,C)$ 
so we can give a qubit interpretation to more vector multiplets for both $N=2$ and $N=4$, at least 
in the case of $SO(4n, C)$ which contains $[SL(2,C)]^{2n}$.

\section{Five-dimensional supergravity}
\la{five}

In five dimensions we might consider:

1) $N=2$ supergravity coupled to $l+1$ vector multiplets where the 
symmetry is $SO(1,1,Z) \times SO(l,1,Z)$ and the black holes carry charges 
belonging to the $(l+1)$ representation (all electric) .

2) $N=4$ supergravity coupled to $m$ vector multiplets where the 
symmetry is $SO(1,1,Z) \times SO(m,5,Z)$ where the black holes carry 
charges belonging to the $(m+5)$ representation (all electric). 

3) $N=8$ supergravity where the symmetry is the non-compact 
exceptional group $E_{6(6)}(Z)$ and the black 
holes carry charges belonging to the fundamental $27$-dimensional 
representation (all electric).

The electrically charged objects are point-like and the magnetic duals 
are one-dimensional, or string-like, transforming according to the 
contrgredient
representation. In all three cases above there exist cubic invariants akin to the 
determinant which yield the corresponding black hole or black string
entropy.

In this section we briefly describe the salient properties of maximal $N=8$ 
case, following \cite{Ferrara:1997ci}. We have 27 abelian gauge
fields which transform in the fundamental representation of
$E_{6(6)}$. The first invariant of $E_{6(6)}$ is the cubic invariant 
\cite{Cartan,Ferrara:1996um,Ferrara:1997ci,Ferrara:1997uz,Andrianopoli:1997hb} 
\be
J_{3} = q_{ij}\Omega^{jl}q_{lm}\Omega^{mn}q_{np}\Omega^{pi}
\ee
where $q_{ij}$ is the charge vector transforming as a $27$ which can be 
represented as traceless $Sp(8)$ matrix.
The entropy of a  black hole with charges $q_{ij}$ is then
given by
\be
S=\pi\sqrt{|J_{3}|}
\la{2/3entropy}
\ee
We will see that a configuration with $J_3 \not = 0$ preserves 1/8 of the supersymmetries.
If $J_3 =0 $ and ${ \partial J_3 \over \partial q^i } \not = 0$  then it preserves
1/4 of the supersymmetries, and finally if ${ \partial J_3 \over 
\partial q^i }=0$
(and the charge vector $q^i$ is non-zero), the configuration preserves
1/2 of the supersymmetries.  We will show this by choosing a particular basis for the 
charges, the general result following by U-duality.

In five dimensions the compact group $H$ is $USp(8)$.
We choose our conventions so that $USp(2)=SU(2)$.
In the commutator of the supersymmetry generators we have
a central charge matrix $Z_{ab}$ 
which can be brought to a normal
form by a $USp(8)$ transformation. In the normal form the central
charge matrix  can be written as
\be
e_{ab} = \pmatrix{ s_1 + s_2 - s_3  & 0& 0& 0\cr 0& 
s_1+s_3 -s_2 &0 &0 \cr
0& 0 & s_2 + s_3 - s_1  & 0 \cr 0 & 0 & 0 & - (s_1 + s_2 + s_3)  } 
\times \pmatrix{ 0 & 1  \cr  -1 & 0}
\ee
we can order $s_i$ so that $s_1 \ge s_2 \ge |s_3|$.
The cubic invariant, in this basis, becomes 
\be
J_3 = s_1 s_2 s_3 
\la{cubic}
\ee
Even though the eigenvalues $s_i$ might depend on the moduli, the
invariant (\ref{cubic}) only depends on the quantized values of the charges.
We can write a generic charge configuration as $U e U^t $,  where $e$ is the normal 
frame as above, and the 
invariant will then be (\ref{cubic}). 
There are three distinct possibilities 
\[
J_3 \neq 0 ~~~~~~~~ s_1,~s_2,~ s_3 \not = 0
\]
\[
J_3 =0 ,~~~{ \partial J_3 \over \partial q^i } \neq 0 ~~~~~~~~ 
s_1,~s_2 \neq 0,~~~~~~~ s_3 =0
\]
\be
J_3 =0 ,~~~~{ \partial J_3 \over  \partial q^i } =0~~~~~~~~~ s_1\neq 0,~~~~~~~s_2, ~ s_3 =0
\la{possib}
\ee
Taking the case of type II on $T^5$ we can choose the rotation in such a way that, for example,   $s_1 $
corresponds to solitonic five-brane charge, $s_2$ to fundamental string winding
charge along some direction and $s_3$ to Kaluza-Klein momentum along the same direction. 
We can see that in this specific example the three possibilities in
(\ref{possib}) break 1/8, 1/4 and 1/2 supersymmetries. The 
respective orbits are
\[
\frac{E_{6}(6)}{F_{4(4)}}
\]
\[
\frac{E_{6}(6)}{SO(5,4) \ltimes T_{16}}
\]
\be
\frac{E_{6}(6)}{SO(5,5) \ltimes T_{16}}
\ee
This also shows that one can generically choose a basis for the
charges so that all others are related by U-duality. 
The basis chosen here is the S-dual of the $D$-brane basis usually chosen for
describing black holes in type II B on $T^5$ . 
All others are related by U-duality to this particular choice.
Note that, in contrast to the four-dimensional case where flipping 
the sign of $J_{4}$ (\ref{Car}) interchanges BPS and non-BPS black holes, the sign of the 
$J_{3}$ (\ref{cubic}) is not important  since it changes under a 
CPT transformation. There is no non-BPS orbit in five dimensions.

In five dimensions there are also string-like configurations
which are the magnetic duals of the  configurations considered here. They 
transform in the contragredient $27'$ representation and the solutions preserving 
1/2, 1/4, 1/8 supersymmetries are characterized in an analogous way. 
We could also have configurations where we  have both point-like  and
string-like ch the point-like charge is uniformly distributed along the string,
it is more natural to consider this configuration as a point-like
object in $D=4$ by dimensional reduction.

It is useful to decompose the U-duality group into the T-duality 
group and the S-duality group. The decomposition reads $E_6 \to
SO(5,5) \times SO(1,1)$,  leading to 
\be
{ 27 } \rightarrow { 16}_{1} + { 10 }_{-2}+ { 1}_{4}
\la{decompts}
\ee
The last term in (\ref{decompts}) corresponds to the NS five-brane charge.
The ${\bf 16}$ correspond to the D-brane charges and the 
${\bf 10 }$ correspond to the 5 directions of KK momentum and
the 5 directions of fundamental string winding, which 
are the charges that explicitly appear in string perturbation 
theory.
The cubic invariant has the decomposition
\be
({ 27})^3 \to {10}_{-2}\ { 10}_{-2}\ { 1}_{4} +
{ 16}_{1}\ { 16}_1\ { 10}_{-2}
\la{cubicdects}
\ee
This is  saying that in order to have
a non-zero area black hole we must have three NS charges
(more precisely some ``perturbative'' charges and a solitonic
five-brane); or  we can have two D-brane charges and one NS charge.
In particular, it is not possible to have a black hole
with a non-zero horizon area with purely D-brane charges.

Notice that the non-compact nature of the groups is crucial in
this classification.
  
\section{$D=5$ black holes and qutrits}
\la{D=5}
So far, all the quantum information analogies involve four-dimensional black holes and qubits.  In order to 
find an analogy with five-dimensional black holes we invoke three state systems called 
{\it qutrits}. 

\subsection{$N=2$ black holes and the bipartite entanglement of two qutrits}
\la{N=2,D=5}
The two qutrit system (Alice and Bob) is described by the state  
\[
|\Psi\rangle ={ a}_{{ A}{ B}}|{ A}{ B}\rangle
\]
where ${A}=0,1,2$, so the Hilbert space has dimension $3^{2}=9$.  
The ${a}_{AB}$ transforms as a $(3,3)$ under $SL(3)_{A} \times SL(3)_{B}$. 
The bipartite  entanglement is measured by the concurrence 
\cite{Fan:2003}
\be
C({ A}{ B})=3^{3/2}|{\rm det}~a_{{ A}{B}}|.
\la{det3}
\ee
The determinant is invariant under $SL(3, C)_{A} \times SL(3, C)_{B}$
and under a duality that interchanges $A$ and $B$. 

The black hole interpretation is provided by $N=2$ supergravity 
coupled to 8 vector multiplets with symmetry $SL(3,C)$ 
where the black hole charges transform as a $9$. The 
entropy is given by
\be
S=\pi|{\rm det}~a_{{ A}{B}}|
\ee

\subsection{$N=8$ black holes and the bipartite entanglement of three qutrits}
\la{N=8,D=5}

As we have seen in section (\ref{five}) in the case of $D=5, N=8$ supergravity, 
the black hole charges belong to the $27$ of $E_{6(6)}$ and the entropy is given 
by (\ref{2/3entropy}).

The qutrit interpretation now relies on the decomposition
\be
E_{6}(C) \supset SL(3,C)_{A} \times SL(3,C)_{B} \times SL(3,C)_{C}
\ee
under which
\be
27 \rightarrow (3,3,1)+ (3',1,3)+(1,3',3')
\la{27}
\ee
suggesting the bipartite entanglement of three qutrits (Alice, Bob, Charlie). 
However, the larger symmetry requires that they undergo at most bipartite 
entanglement of a 
very specific kind, where each person has bipartite entanglement with the 
other two:
\be
|\Psi\rangle = a_{A B}|AB\rangle
             +b^{B}{}_{C}|BC\rangle
              +c^{CA}|CA\rangle
\la{2/3psi}
\ee
where $A=0,1,2$, so the Hilbert space has dimension $3.3^{2}=27$. The three states transforms as a 
pair of triplets under two of the $SL(3)$'s and singlets under the remaining one. Individually, 
therefore, the bipartite entanglement of each of the three states is given by the determinant 
(\ref{det3}). Taken together however, we see from (\ref{27}) that they transform as a complex $27$ 
of $E_{6}(C)$. The entanglement diagram is a triangle with vertices ABC representing the qutrits and 
the lines AB, BC and CA representing the entanglements. See Fig. 3. The N=2 truncation of section 
\ref{N=2,D=5} is represented by just the line AB with endpoints A and B.

\begin{figure}[ht]
\begin{center}
\epsfysize=7cm\epsfbox{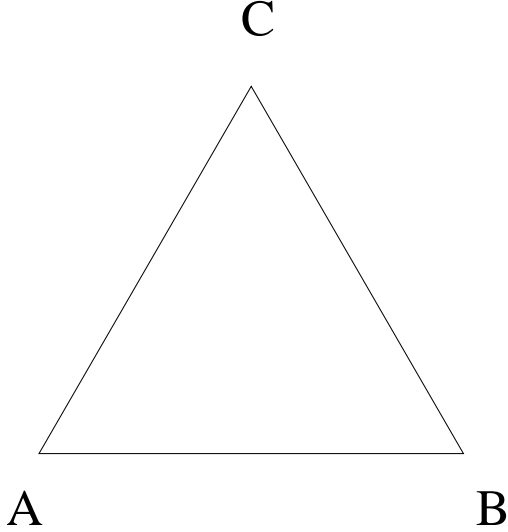}
\end{center}
\caption{ The entanglement diagram is a triangle with vertices ABC 
representing the qutrits and 
the lines AB, BC and CA representing the entanglements.}
\end{figure}

Note that:

1) Any pair of states has an individual in common

2) Each individual is excluded from one out of the three states

The entanglement measure will be given by the concurrence
\be
C(ABC) =3^{3/2}|J_{3}|
\la{2/3tangle}
\ee
$J_{3}$ being the singlet in $27 \times 27 \times 27$:
\be
J_{3} \sim a^{3}+b^{3}+c^{3}+6abc
\la{J3} 
\ee
where the products 
\be
a^{3}=\epsilon^{A_{1}A_{2}A_{3}}\epsilon^{B_{1}B_{2}B_{3}}
{a}_{A_{1}B_{1}}{a}_{A_{2}B_{2}}{a}_{A_{3}B_{3}}
\ee
\be
b^{3}=\epsilon_{B_{1}B_{2}B_{3}}\epsilon^{C_{1}C_{2}C_{3}}
{b}^{B_{1}}{}_{C_{1}}{b}^{B_{2}}{}_{C_{2}}{b}^{B_{3}}{}_{C_{3}}
\ee
\be
c^{3}=\epsilon_{C_{1}C_{2}C_{3}}\epsilon_{A_{1}A_{2}A_{3}}
{c}^{C_{1}A_{1}}{c}^{C_{2}A_{2}}{c}^{C_{3}A_{3}}
\ee
exclude one individual (Charlie, Alice, and Bob respectively), and 
the product
\be
abc={a}_{AB}{b}^{B}{}_{C}{c}^{CA}
\ee
excludes none.

\subsection{Magic supergravities in $D=5$}
\la{magic5}

Just as in four dimensions, one might also consider the ``magic'' supergravities 
\cite{Gunaydin:1984ak,Gunaydin:1983bi,Gunaydin:1983rk}. These 
correspond to the R, C, H, O (real, complex, quaternionic and octonionic) $N=2,D=5$ 
supergravity coupled to $5,8,14$ and $26$ vector multiplets with 
symmetries $SL(3,R),SL(3,C),SU^{*}(6)$ and $E_{6(-26)}$ respectively.
Once again, in all cases there are cubic invariants whose square root 
yields the corresponding black hole entropy \cite{Ferrara:2006yb}. 

Here we demonstrate that the black-hole/qubit correspondence continue to hold for 
these $D=5$ magic supergravities, as well as $D=4$ . Once again, the crucial observation is that, although 
the black hole charges $a_{AB}$ are real (integer) numbers and the 
entropy (\ref{2/3entropy}) is invariant under $E_{6(6)}(Z)$,  the coefficients 
$a_{AB}$ that appear in the wave function (\ref{2/3psi}) are complex. So 
the 2-tangle (\ref{2/3tangle}) is invariant under $E_{6}(C)$ which contains 
both $E_{6(6)}(Z)$ and $E_{6(-26)}(Z)$ as subgroups.  To find a 
supergravity correspondence therefore, we could equally well have chosen  
the magic octonionic $N=2$ supergravity rather than the conventional 
$N=8$ supergravity. The fact that 
\be
E_{6(6)}(Z) \supset [SL(3)(Z)]^{3}
\ee
but
\be
E_{6(-26)}(Z) \not  \supset  [SL(3)(Z)]^{3} 
\ee
is irrelevant. All that matters is that
\be
E_{6}(C) \supset [SL(3)(C)]^{3}
\ee
The same argument holds for the magic real, complex and quaternionic 
$N=2$ supergravities which are, in any case truncations of $N=8$ (in contrast 
to the octonionic). In fact, the example of section 
\ref{N=2,D=5} corresponds to the complex case.

Having made this observation, one may then revisit the conventional 
$N=2$ and $N=4$ cases (1) and (2) of section (\ref{five}).  $SO(l,1)$ is contained in $SO(l+1,C)$ and $SO(m,5)$ is contained in 
$SO(5+m,C)$, so we can give a qutrit interpretation to more vector multiplets 
for both $N=2$ and $N=4$, at least in the case of $SO(6n, C)$ which 
contains $[SL(3,C)]^{n}$.

\section{Conclusions}

We note that the 27-dimensional Hilbert space given in (\ref{27}) 
and (\ref{2/3psi}) is not a subspace of the $3^{3}$-dimensional three 
qutrit Hilbert space given by $(3,3,3)$, but rather a direct sum of 
three $3^{2}$-dimensional Hilbert spaces. It is, however, a subspace 
of the $7^{3}$-dimensional three 7-dit Hilbert space given by $(7,7,7)$. 
Consider the decomposition
\[
SL(7)_{A} \times SL(7)_{B}  \times SL(7)_{C} \rightarrow
SL(3)_{A} \times SL(3)_{B}  \times SL(3)_{C}
\]
under which
\[
(7,7,7) \rightarrow
\]
\[ 
 (3',3',3')+(3',3',3)+(3',3,3')+(3,3',3')+(3',3,3)+(3,3',3)+(3,3,3')+(3,3,3)
\]
\[
+(3',3',1)+(3',1, 3')+(1,3',3')+(3',1,3)+(3',3,1)+(1,3,3')
\]
\[
+(3,3,1)  +(3,1,3)   +(1,3,3)  +(3,1,3')+(3,3',1)+(1,3',3)
\]
\[
+(3',1,1)+(1,3',1)+(1,1,3')+(3,1,1) + (1,3,1)+(1,1,3)
\]
\[
+(1,1,1)
\]
This contains the subspace that describes the bipartite entanglement of 
three qutrits, namely
\[
(3',3,1)+(3,1,3)+(1,3',3')
\]
So the triangle entanglement we have described fits within 
conventional quantum information theory.

Our analogy between black holes and quantum information remains, for the 
moment, just that. We know of no physics connecting them.

Nevertheless, just as the exceptional group $E_{7}$ describes the tripartite 
entanglement of seven qubits \cite{Duff:2006ue,Levay:2006pt}, we have 
seen is this paper that the exceptional group $E_{6}$ describes the 
bipartite entanglement of three qutrits. In the $E_{7}$ case, the 
quartic Cartan invariant provides both the measure of entanglement 
and the entropy of the four-dimensional $N=8$ black hole, whereas in 
the $E_{6}$ case, the cubic Cartan invariant provides both the measure 
of entanglement and the entropy of the five-dimensional $N=8$ black hole.

Moreover, we have seen that similar analogies exist not only for the $N=4$ and $N=2$ 
truncations, but also for the magic $N=2$ supergravities in both four 
and five dimensions (In the four-dimensional case, this had previously 
been conjectured by Levay\cite{Duff:2006ue,Levay:2006pt}).  Murat Gunaydin has 
suggested (private communication) that the appearance of octonions  
implies a connection to quaternionic and/or octonionic quantum mechanics. This was not 
apparent (at least to us) in the four-dimensional $N=8$ case \cite{Duff:2006ue}, but the 
appearance in the five dimensional magic $N=2$ case of $SL(3,R)$, 
$SL(3,C)$, $SL(3,H)$ and $SL(3,O)$ is more suggestive.

\section{Acknowledgements}

MJD has enjoyed useful conversations with Leron Borsten, Hajar 
Ebrahim, Chris Hull, Martin Plenio and Tony Sudbery.  This work was supported 
in part by the National Science Foundation under grant number PHY-0245337 and PHY-0555605. 
Any opinions, findings and conclusions or recommendations expressed in this 
material are those of the authors and do not necessarily reflect the views of 
the National Science Foundation. The work of S.F. has been supported in 
part by the European Community Human Potential Program under contract 
MRTN-CT-2004-005104 Ó Constituents, fundamental forces and symmetries 
of the universeÓ, in association with INFN Frascati National Laboratories 
and by the D.O.E grant DE-FG03-91ER40662, Task C. The work of MJD is supported 
in part by PPARC under rolling grant \uppercase{PPA/G/O}/2002/00474, PP/D50744X/1.


\end{document}